\documentclass[twocolumn, aps, pra, superscriptaddress]{revtex4-1}
\pdfoutput=1
\usepackage[utf8]{inputenc}
\usepackage[english]{babel}
\usepackage[T1]{fontenc}
\usepackage{hyperref}
\usepackage{graphicx}
\usepackage{amsfonts, amsmath, amsthm, amssymb} 
\usepackage{braket}
\usepackage{float}
\usepackage{amsthm}
\usepackage[all]{nowidow}

\begin{document}

\title{Mutually unbiased bases as a commuting polynomial optimisation problem}

\author{Luke Mortimer}
\affiliation{ICFO-Institut de Ciencies Fotoniques, The Barcelona Institute of Science and Technology, 08860 Castelldefels, Spain}

\date{\today}

\begin{abstract}
\noindent
We consider the problem of mutually unbiased bases as a polynomial optimization problem over the reals. We heavily reduce it using known symmetries before exploring it using two methods, combining a number of optimization techniques. The first of these is a search for bases using Lagrange-multipliers that converges rapidly in case of MUB existence, whilst the second combines a hierarchy of semidefinite programs with branch-and-bound techniques to perform a global search. We demonstrate that such an algorithm would eventually solve the open question regarding dimension 6 with finite memory, although it still remains intractable. We explore the idea that to show the inexistence of bases, it suffices to search for orthonormal vector sets of certain smaller sizes, rather than full bases. We use our two methods to conjecture the minimum set sizes required to show infeasibility, proving it for dimensions 3. The fact that such sub-problems seem to also be infeasible heavily reduces the number of variables, by 66\% in the case of the open problem, potentially providing an large speedup for other algorithms and bringing them into the realm of tractability.
\end{abstract}

\maketitle

\section{Introduction to mutually unbiased bases}

A set of vectors form a complete orthonormal basis if each has a magnitude of one, have pairwise inner products equal to zero, and completely span the space in which they are defined. For example, the following is a complete orthonormal basis in $\mathbb{C}^2$, generally referred to as the computational basis:
\begin{align}
     v_{11} = (1 ~~ 0) \qquad v_{12} = (0 ~~ 1)
\end{align}

Two orthonormal bases are considered mutually unbiased if the inner products between their vectors each has a magnitude of $\frac{1}{\sqrt{d}}$. This extends to sets of more than just two, such that each basis is mutually unbiased with respect to every other basis, for example the following set of 3 mutually unbiased bases (MUBs) in dimension 2:
\begin{align}
    v_{11} &= (1 ~~ 0) ~~~~~~~~~~~ v_{12} = (0 ~~ 1) \\
    v_{21} &= \frac{1}{\sqrt{2}}(1 ~~ 1) \qquad v_{22} = \frac{1}{\sqrt{2}}(1 ~~ -1) \\
    v_{31} &= \frac{1}{\sqrt{2}}(1 ~~ i) \qquad v_{32} = \frac{1}{\sqrt{2}}(1 ~~ -i)
\end{align}

Such bases have significant use in the field of quantum information theory, providing measurements maximising Bell inequalities \cite{tabia2022bell,tavakoli2021mutually}, their use in quantum state tomography \cite{rebon2013phase,yuan2016quantum}, or their use in secure quantum communication protocols \cite{mafu2013higher}. This is due to the property that preparing in one basis and measuring in a mutually unbiased basis results in a uniform probability of any outcome, making it difficult for an attacker to gain information. 

In any dimension $d$ it is known that there are at most $d+1$ MUBs. More specifically, in a given dimension $d$ that is the power of a prime it is known that one can always find the maximal set of $d+1$ MUBs \cite{wootters1987wigner}. It remains a decades-old open question exactly how many MUBs exist in dimensions that are not the power of a prime, the lowest being dimension 6. There exists a conjecture, known as Zauner's conjecture, stating that one can find at most 3 MUBs in dimension 6 \cite{zauner1999grundzuge}. This claim is supported by numerical searches which have failed to find 4 MUBs in dimension 6 despite extensive effort \cite{colomer2022three,raynal2011mutually}.

\section{The MUB problem as polynomial optimisation}

One of the most well-studied forms of problem is that of polynomial optimisation, which in the general form consists of trying to minimize a real polynomial $f(\vec{x})$ over a vector $\vec{x}$, subject to equality constraints $g(\vec{x}) = 0$ and positivity constraints $h(\vec{x}) > 0$. These polynomials can (and in this work will be) non-convex and non-linear, thus allowing all problems in NP to be expressed \cite{jacquet2020reformulation}. If the polynomials were all convex, this type of problem could be solved easily using convex optimisation techniques, many of which are known to converge to finite precision with polynomial time complexity \cite{karmarkar1984new,laurent2005semidefinite}. The idea of converting the MUB existence problem into a polynomial optimisation problem has been studied before in the non-commuting complex case \cite{gribling2021mutually} or in the case of studying MUB constellations \cite{brierley2010mutually}.

Here an objective function $f(\vec{x})$ is unnecessary, since we are only trying to find any valid set of MUBs, rather the set of MUBs which minimizes some function. The same is true for the positivity constraints, since in the MUB problem we only need to constrain that various magnitudes of inner products equal either 0, 1, or $\frac{1}{\sqrt{d}}$. So, for the problem of finding $n$ MUBs in dimension $d$, we begin with the problem of finding complex vectors $\vec{v}_{ij}$, where here the first index indicates the basis (from $1 \to n$) and the second indicates the vector within the basis (from $1 \to d$):
\begin{align}
    \vec{v}_{ii} ^ \dagger \vec{v}_{ii} &= 1  \\
    \vec{v}_{ik} ^ \dagger \vec{v}_{il} &= 0   \\
    |\vec{v}_{ik} ^ \dagger \vec{v}_{jl}| &= \frac{1}{\sqrt{d}}   \label{eqn:mag} \\
    i \ne j &\qquad k \ne l
\end{align}

In order to simplify Equation \ref{eqn:mag} we introduce two new real variables $c_t,d_t$ for each of the pairs, constrained to the circle of magnitude $\frac{1}{\sqrt{d}}$. This allows us to keep the equations second-order, versus removing the norm by squaring the constraint which would allow for less variables with the trade-off of a fourth-order equation set. Testing both version we found it is significantly better regarding both the memory and the time required for both of our algorithms if the order is kept as low as possible. We also replace each element of each vector with a combination of the real and complex parts, such that element $m$ of $\vec{v}_{ij}$ becomes $v_{iim} = a_{iim}+ib_{iim}$, giving the following polynomial equation set over the reals:
\begin{align}
    \sum_{m=1}^d (a_{iim}-ib_{iim}) (a_{iim}+ib_{iim}) &= 1 \\
    \sum_{m=1}^d (a_{ikm}-ib_{ikm}) (a_{ilm}+ib_{ilm}) &= 0 \\
    \sum_{m=1}^d (a_{ikm}-ib_{ikm}) (a_{jlm}+ib_{jlm}) &= c_{i+dj}+id_{i+dj} \\
    c_{i+dj}^2 + d_{i+dj}^2 &= \frac{1}{d} \\
    i \ne j \qquad &k \ne l
\end{align}

One can then expand the products, equating the real and complex sides of each equation, as well as combining all variables into a single vector for computational simplicity. Now we have a set of real equations in which every feasible point is a set of MUBs, thus proving infeasibility of the equation set for dimension $d$ and number of bases $n$ proves the inability to generate MUBs for that $d$ and $n$. The number of variables quickly becomes large, as shown in Table \ref{tbl:varNums1}, which becomes a problem when discussing the global optimisation required to prove infeasibility, .

\begin{table}[h!]
\centering
\begin{tabular}{|c | c c c c c|} 
 \hline
 n/d & 2 & 3 & 4 & 5 & 6 \\
 \hline
 2 & 12 & 27 & 48 & 75 & 108 \\ 
 3 & 24 & 54 & 96 & 150 & 216 \\
 4 & 40 & 90 & 160 & 250 & 360 \\
 5 & - & 135 & 240 & 375 & - \\
 6 & - & - & 336 & 525 & - \\
 7 & - & - & - & 700 & - \\
 \hline
\end{tabular}
\caption{Number of variables when converting the mutually unbiased basis problem into a real polynomial optimisation problem, without any reductions. The formula for this is $nd^2+\frac{1}{2}d^2n(n-1)$. Note that due to the non-convexity of the problem the worst-case complexity of such general polynomial optimisation is exponential in the number of variables, thus here the open question of $d=6$ and $n=4$ remains highly intractable.}
\label{tbl:varNums1}
\end{table}

\section{Reducing the number of variables using symmetries}

Whilst one could use the algorithms presented in this work to optimise the polynomial problem in its current state, we were unable to perform global optimisation for even small sets in dimensions 2 and 3. Luckily for us, MUBs have a large number of rotational and reflectional symmetries that we can exploit to significantly reduce the number of variables and equations.

The first of these reductions can be derived from the fact that a unitary rotation applied globally to a set of MUBs preserves the mutually unbiasedness \cite{romero2005structure}. As such, one can always choose the first basis to be any valid orthonormal basis without loss of generality. For the sake of simplicity, we choose the computational basis mentioned earlier, which as well as removing a large number of variables from the optimisation also allows for a further simplification: given that all inner products must have a fixed inner product and given that each vector in the computational basis only affects one element of each vector, one can now explicitly constraint the magnitude of every element of every other vector to be $\frac{1}{\sqrt{d}}$, for example in dimension 2:
\begin{align}
    \text{if  } \quad \vec{v}_{11} &= (1 ~~ 0) \quad \text{and} \quad \vec{v}_{12} = (0 ~~ 1) \\
    \implies \vec{v}_{21} &= \frac{1}{\sqrt{2}}(e^{i\theta_1} ~~ e^{i\theta_2})
\end{align}

One can also choose to fix some of the rotational symmetries by fixing the first vector of the second basis, for instance to the uniform vector of $\vec{v}_{21} = \frac{1}{\sqrt{d}}(1 ~~ 1 ~~...~~ 1)$. A similar argument can be made for fixing the first element of every vector to be $\frac{1}{\sqrt{d}}$, further reducing the number of variables we need to optimise. These can both be done without loss of generality due to the inner product being unaffected by the phases of the bases. If one could find 4 MUBs in dimension 6 without such restrictions, one could then simply add a phase to each basis/vector until we have MUBs that follow the aforementioned restrictions.

Continuing with the idea of symmetry reductions, one can consider a number of other symmetries which preserve the various inner products of the problem, specifically certain permutations of variable. For instance, given a set of MUBs, swapping the first and second bases is guaranteed to preserve the mutually unbiasedness. The same can be said about swapping two vectors within a basis, swapping the first and second elements of all vectors, as well as taking the conjugate of everything. All of these are transformations which preserve the inner product. In order to reduce the size of the search space we define positivity constraints for each of the aforementioned symmetries, such that one can always swap the sign of the constraint function by performing the corresponding transformation on the equation set, whilst retaining mutual unbiasedness.

For example, since it is known that we can swap vectors 1 and 2 in the 3rd basis without loss of generality, once can define some function of the form $h_1(\vec{v}_{31},\vec{v}_{32}) = \sum_{m=1}^d ( a_{31m} + b_{31m} )  - \sum_{m=1}^d ( a_{32m} + b_{32m} )$, the difference between the sums of the components of each vector. This can always be made positive by MUBs, since if a set of MUBs was to make this negative one could simply swap the two bases, resulting in the negative of the function. A similar function can be defined for the conjugation symmetry: $h_2(\vec{b}) = \sum_m b_m$. Given MUBs making this function negative one can take the conjugate of the MUBs resulting in the negative of all of the imaginary components $b_m$, thus flipping the sign of this function whilst retaining the mutual unbiasedness. Adding each of these functions as positivity constraints to the polynomial problem reduces the size of the feasible set without loss of generality in the infeasibility case.

\section{Reducing the number of variables using sub-bases}

Whilst we have reduced the number of variables significantly with the reductions so far, we are still considering the full equation set. If we now shift the focus to not finding full sets of MUBs, but instead showing that MUBs are infeasible, the question can be asked, is any subset of the equations infeasible? If this were the case the infeasibility would extend to the full equation set, whilst being much simpler to prove due to the reduction in number of equations. Here we focus on the idea of optimising sub-bases, which we define as a series of orthonormal vectors which does not contain enough vectors to fully span the space, thus making it a subset of some basis. For example, in dimension 3 one could consider the sub-basis of the computational basis:
\begin{align}
     v_{11} = (1 ~~ 0 ~~ 0) \qquad v_{12} = (0 ~~ 1 ~~ 0)
\end{align}

Since finding a valid sub-basis is a requirement for finding a valid basis, if a given problem does not permit sub-bases, it certainly doesn't permit full bases. As such we extend the notation with the mutually unbiased property, turning our search from mutually unbiased bases (MUBs) to the sub-problem of searching for mutually unbiased sub-bases (MUSBs), which we denote by listing the number of vectors per set. For example in dimension 3, the MUSB problem with sizes $\{3,2,1\}$ would imply the search for a full basis (of 3 vectors), a sub-basis of 2 vectors, plus a sub-basis of 1 vector, each being mutually unbiased with each other. The choice of such sub-basis sizes, in order to minimize the size of the problem whilst still remaining infeasible in the cases that should be infeasible, is something that was determined by starting with the smallest MUSB problem for a given $d$ and $n$, then increasing the size until the problem appears to become infeasible as decided by the upcoming methods. Combining all reductions results in a significant drop in the number of variables, as given by table \ref{tbl:varNums2}. For the open question of 4 MUBs in dimension 6 this results in a 66\% reduction ($360 \to 122$).

\begin{table}[h!]
\centering
\begin{tabular}{|c | c c c c c|} 
 \hline
 n/d & 2 & 3 & 4 & 5 & 6 \\
 \hline
 2 & 0          & 0           & 6           & 8           & 50 \\ 
 3 & 2          & 4           & 14          & 28          & 119 \\
 4 & 6          & 10          & 24          & 42          & 122 \\
 5 & -          & 18          & 36          & 58          & - \\
 6 & -          & -           & 50          & 76          & - \\
 7 & -          & -           & -           & 96          & - \\
 \hline
\end{tabular}
\caption{Number of variables when converting the mutually unbiased basis problem into a real polynomial optimisation problem, with all reductions. Here the set sizes used are those with the lowest number of variables whilst still retaining infeasibility in the presumed infeasible cases.}
\label{tbl:varNums2}
\end{table}

\section{Method 1: Search Space Without Local Minima}

Given this reduced set of equations, we first considered methods to ensure that this correctly conveys the problem, finding MUBs in the cases where they should exist and failing to find MUBs in those where they are not. To do this we use the idea of combining our many second-order equality constraints into one giant fourth-order objective function, which could then be minimized without constraints. Consider first some general equation set in which we want to find a common root:
\begin{align}
    h_i(\vec{x}) = 0 \qquad \forall i
\end{align}

which we can then combine by squaring each equation and summing:
\begin{align}
    f(\vec{x}) = \sum_i h_i^2(\vec{x}) = 0
\end{align}

Since each term has to be non-negative, the only way such a function can reach zero is if all of the squared equations also reach zero, thus transforming the problem into an unconstrained optimisation. Here we just use the equality constraints of the reduced equation set and ignore the positivity constraints used for enforce some symmetries, since these only provide a slight improvement to this method but add quite a bit of complexity. Our aim with this optimisation is to converge rapidly to MUSBs if they exist, with a lack of convergence serving as a suggestion that such a set of MUSBs (and thus a corresponding set of MUBs) is unlikely to exist.

Based on the method of Lagrange multipliers, we begin by integrating our polynomial $f(\vec{x})$ by one of our variables, $x_m$, specifically one that is known to be able to reach zero without loss of generality. In the case of the MUB problem this is fine, since we have a number of variables already assumed to be zero due to rotational symmetries, however equally one could create a new variable and integrate by it to create something that resembles a traditional Lagrange multiplier. Regardless of the choice of variable, the result is then a new fifth-order polynomial $f_m(\vec{x})$ with a number of interesting properties. Considering the partial derivatives of this function with respect to every variable, we find that one of these derivatives is our original function, whilst every other derivative has our integrating variable $x_m$ as a factor. As an arbitrary example, if our original function was:
\begin{align}
    f(\vec{x}) = x_0^2 + x_0x_1 + x_2^4
\end{align}
and we choose to integrate by $x_1$:
\begin{align}
    f_1(\vec{x}) = \int f(\vec{x}) ~ dx_1 = x_1x_0^2 + \frac{1}{2}x_0x_1^2 + x_1x_2^4
\end{align}
which has partial derivatives:
\begin{align}
    \frac{\partial}{\partial x_0} f_1(\vec{x}) &= 2x_1x_0 + \frac{1}{2}x_1^2 \\
    \frac{\partial}{\partial x_1} f_1(\vec{x}) &= f(\vec{x}) \\
    \frac{\partial}{\partial x_2} f_1(\vec{x}) &= 4x_1x_2^3 \\
\end{align}
From this it can be seen that the set of partial derivatives can only be all zero, a stationary point, if our original function is zero. The other derivatives being zero can be guaranteed to exist if the variable $x_1$ can also go to zero without loss of generality. As such, we have created a search space in which every single stationary point is a solution to our problem. In the case of MUBs this means a search space in which all stationary points are MUBs, with no stationary points existing in the cases in which MUBs do not exist.

Now the task is to find a stationary point, which we do using a Newton-style iterative approach. This is done by calculating the matrix of second-derivatives $H$ as well as the vector of partial derivatives $\vec{g}$ of our function $f_m(\vec{x})$, then solving the system of linear equations $H\vec{p}=\vec{g}$ to get the update direction $\vec{p}$. We then follow this direction, scaled by some factor $0<\alpha\le 1$. The result in an algorithm which will eventually converge (given a small enough step size) if there are MUBs, whilst endlessly traversing the search space if MUBs do not exist. As such it does not constitute a formal proof of non-existence, simply serving as suggestion as to whether the given problem is feasible or not.

For the numerical computation of this method we use the library Eigen \cite{eigen} for the linear system solving required for the Newton step. We tested a number of solving methods but found that a partial-pivot LU decomposition works the best in our case, offering an effective blend of speed and accuracy, as well as offering a parallel implementation. Whilst we did experiment with a line-search for choosing the parameter $\alpha$, this seemed to require more time than it saved. We also tested a quasi-Newton update method in which we did not explicitly calculate the full Hessian, but the loss of accuracy resulted in very slow convergence. We also add a small ($10^{-10}$) term to the diagonals of the Hessian to improve the numerical stability.

\section{Method 2: Branch and bound combined with a hierarchy of SDPs}

For our second method, we use another optimization technique: a hierarchy of semidefinite programming relaxations similar to the Lasserre \cite{lasserre2001global} or NPA \cite{navascues2008convergent} hierarchies for commuting and non-commuting polynomial optimization, respectively. The idea behind such techniques is to relax the problem and then add constraints of increasing complexity (referred to as levels in some hierarchy) to tighten the relaxation, often with guarantees of finite convergence. We begin by taking our non-linear equations and creating new variables to linearize them, turning each term $x_ix_j$ into a new variable $x_{ij}$, for example:
\begin{align}
    x_1x_2 + x_2^2 + x_1  \to x_{12} + x_{22} + x_1
\end{align}

We start by assuming that the new variables have the same bounds as the originals, meaning at this point the relaxation is quite bad, for instance $x_{ij}$ could be negative despite $x_i$ and $x_j$ being positive since for now there is no constraint linking them. It is, however, fast to compute, since we only have linear equations and thus it has become a convex optimisation problem. To then begin to constrain $x_{ij}$ to be closer to $x_ix_j$ we add a hierarchy of positive-semidefinite (PSD) constraints, which are also convex and thus the relaxation remains convex. The first level of such a hierarchy considers the matrix whose first row contains all first-order monomials:
\begin{align}
    \begin{pmatrix}
        1 & x_1 & x_2 \\
        x_1 & x_{11} & x_{12} \\
        x_2 & x_{12} & x_{22}
    \end{pmatrix} \succcurlyeq 0
\end{align}

Going to the second level in our two variable example would look as follows, with second order terms on the top row and then forming the corresponding products:
\begin{align}
    \begin{pmatrix}
        1 & x_1 & x_2 & x_{11} & x_{12} & x_{22} \\
        x_1 & x_{11} & x_{12}  & x_{111}  & x_{112}  & x_{122} \\
        x_2 & x_{12} & x_{22}  & x_{112}  & x_{122}  & x_{222} \\
        x_{11} & x_{111} & x_{112}  & x_{1111}  & x_{1112}  & x_{1122} \\
        x_{12} & x_{112} & x_{122}  & x_{1112}  & x_{1122}  & x_{1222} \\
        x_{22} & x_{122} & x_{222}  & x_{1122}  & x_{1222}  & x_{2222} \\
    \end{pmatrix} \succcurlyeq 0
\end{align}

What we now have is an optimisation over a much larger set of variables with linear equality constraints corresponding to the MUB problem and linear inequality constraints to break symmetries, all to some chosen level of semidefinite hierarchy with the hope that at some level this convex relaxation of the problem will become infeasible in the cases of MUB non-existence. However, we find that even high levels (>5) of this hierarchy alone do not become infeasible, even for small cases. Some works dealing with the NPA hierarchy suggest that one may need at least level 12 of that hierarchy \cite{gribling2021mutually}. If the level needed for the commuting case is similar, it would require solving a semidefinite program of $110^{12}$ variables to solve the $d=6$ $n=4$ case, which is at least six orders of magnitude larger than what modern supercomputers can solve within a reasonable time. As such, we ask the question: maybe we don't need to solve it in a single convex optimisation? Perhaps splitting it into a larger number of smaller convex regions would be sufficient?

Consider now the first level of the first-order relaxation of a single variable. Since it's a 2x2 matrix, there are only 2 eigenvalues. For both to be positive, the product of eigenvalues needs to be positive and thus the determinant needs to be positive:
\begin{align}
    \begin{pmatrix}
        1 & x_1 \\
        x_1 & x_{11}
    \end{pmatrix} \succcurlyeq 0 \qquad \implies \qquad x_{11} - x_1^2 \ge 0
\end{align}

\begin{figure}[ht]
\centering
\includegraphics[width=\columnwidth]{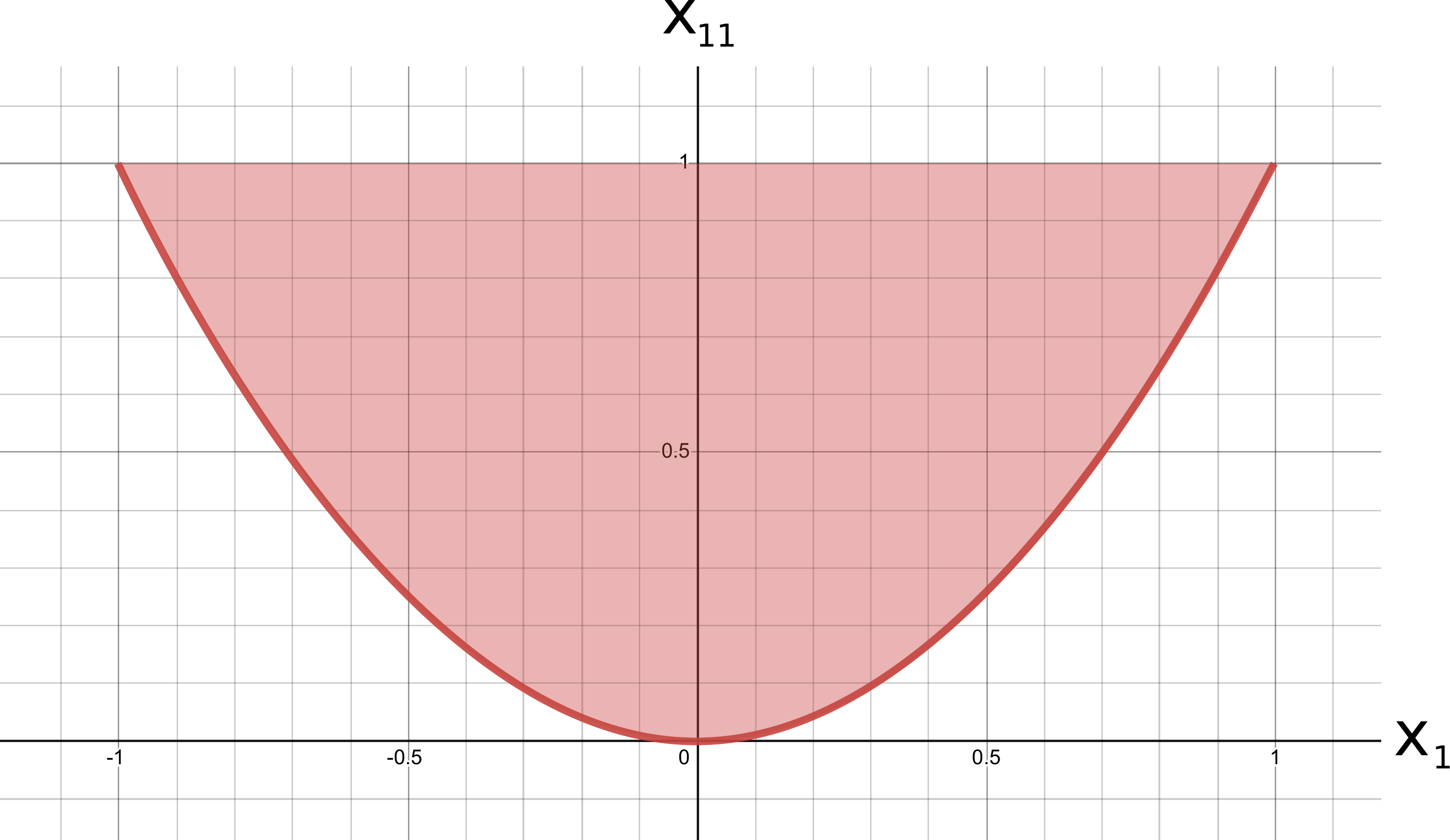}
\caption{The search space created from the constraint $x_{11} - x_1^2 \ge 0$ as well as the box constraint $-1 \le x_1 \le 1$}
\label{fig:xxCone}
\end{figure}

Assuming we also use box constraints of $-1<x_1<1$, which we can assume in the MUB case since all of the magnitudes are bounded, the search space looks as given in Figure \ref{fig:xxCone}.

\begin{figure}[ht]
\centering
\includegraphics[width=\columnwidth]{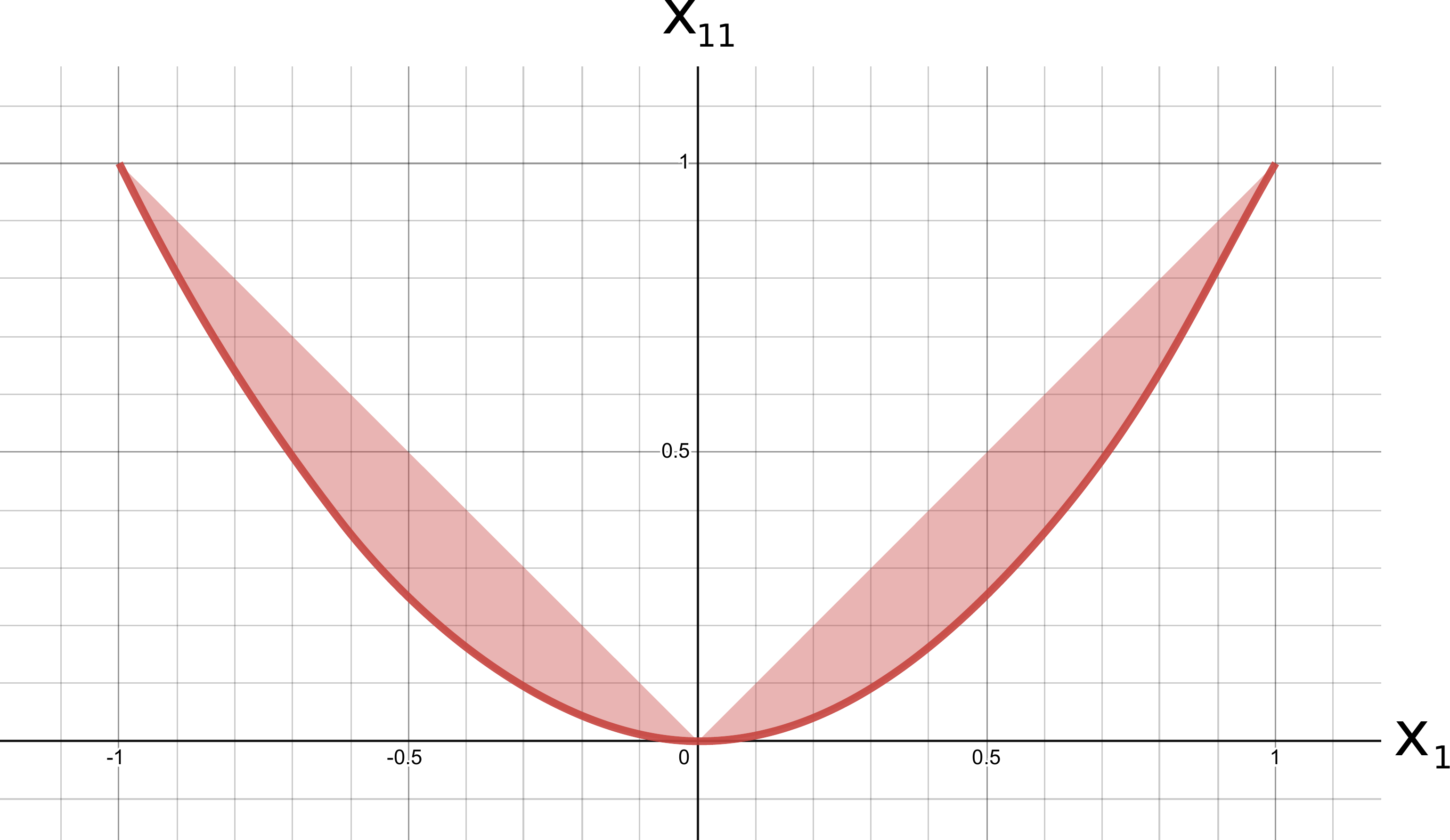}
\caption{The two convex search spaces created from the constraint $x_{11} - x_1^2 > 0$ with the box constraints $-1 \le x_1 \le 0$ and $0 \le x_1 \le 1$}
\label{fig:xxConeSplit}
\end{figure}

As one can see, this constraint still leaves many points such as $(0,1)$ well within the search space whilst being very far from the $x_{11}=x_1^2$ line to which we want to constrain. This shows the issue with trying to convexify such an equality constraint. Our approach here is to stay in the first level of the hierarchy whilst splitting such a convex region into two. Whilst for many styles of constraint splitting a region into two would only half the size of the region, in our case we can constrain further since we know the edge we are trying to converge towards, as shown by Figure \ref{fig:xxConeSplit}.

By first performing the first-level relaxation without any branching we obtain some point in the search space. Checking the error of each variable versus its square $\text{err}(x_i) = (x_{ii}-x_i^2)^2$ we get a heuristic for which variables need constraining tighter. We found that a good branching method is to choose the variable with the highest error, splitting it in two (as in Figure \ref{fig:xxConeSplit}) at the feasible point that was found by the previous iteration. As such we have divided our whole search space into two, whilst also constraining tighter towards the set of valid first and second-order monomials. One can also choose to add an objective function to ensure that we find the "most-feasible" point, done by adding a new variable $\lambda$ and asking that the moment matrix plus $\lambda$ times the identity is positive-semidefinite, then minimizing $\lambda$.

By repeating this process one can see that we will eventually reach regions which are sufficiently small that the maximum error for any variable becomes less than some $\epsilon$. In the MUB non-existence cases, under the assumption that there exists some $\epsilon$ such that one cannot find MUBs even allowing such error in the relations (e.g. that the various inner products are now constrained within $\frac{1}{\sqrt{d}} \pm \epsilon$), such small regions will result in convex infeasibility, which can be efficiently detected. By checking that a large number of these convex regions are all infeasible one can show that the original equation set is infeasible, thus proving the MUB inexistence case.

Whilst in the above case we consider the first-level relaxation, moving to the second level must be at least as constraining as the first, since the second-level PSD matrix contains the first as a submatrix. The procedure works the same as we move to higher-level relaxations, except now we have higher-order terms in the first row of our PSD matrix, making it much larger. For example, given 18 variables (as in the case of our reduction of the 5 MUBs in dimension 3 case), the first-level relaxation has a PSD matrix of 19x19, the second a matrix of 343x343, the third 6175x6175 and so on. Given that SDPs in the size of 300x300 are already difficult for desktop computers, doing a large number of 6000x6000 SDPs is near intractable. The hope, however, would be that by doing a tighter relaxation one does not need to perform so many branches to prove full infeasibility, giving a trade-off between the number of iterations needed and the level of the relaxation. 

For the numerical computation of this method we have tested two SDP solvers: MOSEK \cite{mosek} and Splitting Cone Solver (SCS) \cite{scs}. For the small cases and the first-level SDPs MOSEK works well, but has an incredibly high memory profile due to its need to store various second-order derivatives, however this allows it high accuracy and fast (logarithmic) convergence versus other methods. The alternative is SCS, which as a first order method has a very minimal memory usage, often using 100x less memory than MOSEK, at the cost of significantly longer convergence times. For all second-level optimisation we use SCS, since otherwise memory quickly becomes an issue, with MOSEK using around 80 GB for proving dimension 3 non-existence whilst SCS uses only 100 MB. With both libraries we exploit their sparsity and model parameterization features.

\section{Results}

We first begin by showing the amount of MUBs that seem to be feasible and infeasible using our first method, shown in Table \ref{tbl:resultsFeasible}, given as the number of iterations it required before we either found a satisfying vector or we stopped the optimization. With this we show how easily the algorithm finds solutions to contrast with the amount of time spend on the presumed non-existence cases.

\begin{table}
\centering
\begin{tabular}{|c | c c c c c c |} 
 \hline
 $n$ / $d$ & 2 & 3 & 4 & 5 & 6 & 10\\
 \hline
 2 & 1      & 1      & 55     & 200    & 84     & 175    \\ 
 3 & 59      & 55    & 61     & 63     & 73     & 226     \\
 4 & 100000 & 113    & 75     & 104    & 100000 & 100000  \\
 5 & 100000 & 100000 & 87     & 96     & 100000 & 100000  \\
 6 & 100000 & 100000 & 100000 & 235    & 100000 & 100000 \\
 7 & 100000 & 100000 & 100000 & 100000 & 100000 & 100000 \\
 \hline
\end{tabular}
\caption{The number of iterations required to either find the MUSB or until reaching the chosen iteration limit. These results, if we assume reaching the limit of 100000 iterations implies infeasibility, agree with known results and Zauner's conjecture. Here the criteria for convergence was reaching a value below $10^{-13}$ in the combined equation, roughly corresponding to a maximum error of $10^{-7}$ in any of the original equality constraints.}
\label{tbl:resultsFeasible}
\end{table}

The algorithm in general seems to find MUSBs with set sizes with less deviation (i.e. $\{4,4,3,3\}$) faster than those that are heavily deviated (i.e. $\{6,5,2,1\}$), presumably because there are more MUSBs to find due to an increased number of symmetries. For highly uneven set sizes like $\{9,7,1,1,1,1,1,1,1,1\}$ in dimension 9, the algorithm struggles, sometimes taken several thousand iterations. For high dimensions (>15) there is some instability that develops from the large linear systems, to combat this we lower the step size $\alpha$ and in some cases add a larger diagonal term to the Hessian to improve the condition number and thus numerical stability of the solver. 

Now applying the branch-and-bound algorithm, we correctly convergence to the expected results in all cases in dimension 2 and 3. Whilst it was expected that we would not find MUSBs of sizes \{2,1,1,1\} in dimension 2, as far as the authors are aware it has not been shown that one cannot find MUSBs of sizes \{2,1,1,1,1\} in dimension 3. We show in Table \ref{tbl:resultsSDP1} the number of SDPs that were required when branching with the first level of the hierarchy, whilst Table \ref{tbl:resultsSDP2} shows the number using the second level. As expected, significantly less branches were needed as higher levels of the hierarchy were used, however using the third level or higher is intractable for anything higher than dimension 2. 

\begin{table}[h!]
\centering
\begin{tabular}{|c | c c c c c|} 
 \hline
 $n$ / $d$ & 2 & 3 & 4 & 5 & 6 \\
 \hline
 2 & 0   & 0   & 11   & 43   & ? \\ 
 3 & 2   & 40   & 81   & 19279   & ? \\
 4 & \textbf{11}   & 337   & 18004   & ?   & \textbf{?} \\
 5 & -     & \textbf{17153}   & ?   & ?   & - \\
 6 & -     & -     & \textbf{?}   & ?   & - \\
 7 & -     & -     & -     & \textbf{?}   & - \\
 \hline
\end{tabular}
\caption{Number of SDPs required to reach either convergence to an MUSB or to prove infeasibility when using the branch-and-bound algorithm with the first level of the hierarchy. Convergence to an MUSB is defined as an SDP being feasible with a largest monomial error of at most $10^{-8}$. Bold text indicates the relevant problems for proving infeasibility. Here the 17153 iterations for dimension 3 takes around 3-4 minutes on a standard desktop.}
\label{tbl:resultsSDP1}
\end{table}

\begin{table}[h!]
\centering
\begin{tabular}{|c | c c c c c|} 
 \hline
 $n$ / $d$ & 2 & 3 & 4 & 5 & 6 \\
 \hline
 2 & 0           & 0   & 13   & 13   & ? \\ 
 3 & 1           & 18   & 35   & ?   & ? \\
 4 & \textbf{1}  & 22   & ?   & ?   & \textbf{?} \\
 5 & -           & \textbf{7}   & ?   & ?   & - \\
 6 & -           & -     & \textbf{?}   & ?   & - \\
 7 & -           & -     & -     & \textbf{?}   & - \\
 \hline
\end{tabular}
\caption{Number of SDPs required to reach either convergence to MUSB or proving infeasibility when using the branch-and-bound algorithm with the second level of the hierarchy. Bold text indicates the relevant problems for proving infeasibility. Here the 7 iterations for dimension 3 takes around a minute on a standard desktop with SCS.}
\label{tbl:resultsSDP2}
\end{table}

As this algorithm runs, it regularly declares regions as infeasible, each of which has an associated area. Since in our problem the total search space is bounded (each variable between $\pm\frac{1}{\sqrt{d}}$) we can obtain the fraction of the search space that has been declared infeasible. If the entire problem is infeasible then this total value increases until reaching 100\%, otherwise a solution is found before then. By considering the time taken to reach the current percentage of infeasibility and extrapolating to 100\%, the algorithm progressively outputs an estimate for the time it will take to prove infeasibility. For the $d=3$ and $n=5$ case (18 variables) this states 5 minutes at the start, the algorithm then finishes in around 4. When running for $d=4$ and $n=6$ (50 variables) it estimates several months. For $d=6$ and $n=4$ (118 variables), the open problem, it estimates something on the order of hundreds of millions of years. Even assuming perfect parallelisation and running the algorithm on the worlds best supercomputer (at the time of writing the American ``Frontier'' \cite{rajaraman2023frontier}) it would still take hundreds of years to solve this open problem with this algorithm. As such, further reductions are needed, otherwise we may simply have to wait until classical computational power reaches a level such that this becomes tractable.

Considering the results from both methods, we notice there appears to exist some sort of pattern in the minimum sizes that are infeasible, for instance we find that for dimension 6 $\{6,3,3,3\}$ does not seem to be feasible, whilst $\{6,3,3,2\}$ is feasible and the algorithm finds a solution in very few iterations. Although our first thoughts were that perhaps it relates to the total number of vectors, it cannot be so simple due to the lack of freedom of the final vector in an orthonormal set, for instance that in dimension 3, $\{2,2,2,2,2\}$ and $\{3,3,3,3,3\}$ will both be equally infeasible despite having very different total vector counts. We list a selection of feasible and infeasible sets for dimensions 3, 4, 5 and 6 in Tables \ref{tbl:resultsSizes3}, \ref{tbl:resultsSizes4}, \ref{tbl:resultsSizes5} and \ref{tbl:resultsSizes6} respectively.

One possible conjecture is that the MUSB problem is infeasible if the number of independent constraints is at least the number of variables minus one. Whilst there are known theorems for the number of solutions of linear systems that are overdetermined and underdetermined \cite{williams1990overdetermined}, the results do not extend trivially to non-linear systems. Whilst there are some MUSBs such as $\{6,6,1,1\}$ which have more constraints (95) than variables (92) despite being feasible, one can perform an analysis of the equations to show that some are dependant on others. We have found a set of 10 constraints for $\{6,6,1,1\}$ that even when removed, the optimisation seems to always find a solution which satisfies them, suggesting that the other 85 equations were sufficiently constraining. By extension this would also mean these constraints are redundant for $\{6,6,3,1\}$, thus taking it from 149 equations to 139 equations vs 136 variables, leaving it still infeasible under this conjecture and agreeing with the optimisation results.

\begin{table}[ht]
\centering
\begin{tabular}{| c | c c c c |} 
 \hline
 set sizes & vecs & vars & eqns & result \\
 \hline
 \{3,3,3,3\} &  12  & 74 & 101 & proven feasible \\ 
 \hline
\{1,1,1,1,1\} &  5  & 18 & 15 & proven feasible \\ 
\{2,1,1,1,1\} &  6 & 18 & 18 & \textbf{proven infeasible}  \\ 
\{3,1,1,1,1\} &  7  & 18 & 18 & \textbf{proven infeasible}   \\ 
 \hline
\end{tabular}
\caption{A list of some set sizes for dimension 3 as well as the conclusion made using our optimisation techniques. Here ``proven feasible'' means a valid MUSB was found to a tolerance of $10^{-13}$ using method 1 whilst ``proven infeasible'' means that we have branched the entire search space to infeasibility using method 2.}
\label{tbl:resultsSizes3}
\end{table}

\begin{table}[ht]
\centering
\begin{tabular}{| c | c c c c |} 
 \hline
 set sizes & vecs & vars & eqns & result \\
 \hline
 \{4,4,4,4,4\} &  258  & 20 & 354 & proven feasible   \\ 
 \hline
\{1,1,1,1,1,1\} &  6  & 36 & 26 & proven feasible   \\ 
\{2,1,1,1,1,1\} &  7  & 36 & 30 & proven feasible   \\ 
\{2,2,1,1,1,1\} &  8  & 50 & 45 & proven feasible   \\ 
\{2,2,2,1,1,1\} &  9  & 64 & 62 & proven feasible   \\ 
\{2,2,2,2,1,1\} &  10  & 80 & 82 & \textbf{seems infeasible}   \\ 
 \hline
\{4,1,1,1,1,1\} &  9  & 36 & 34 & proven feasible   \\ 
\{4,2,1,1,1,1\}  &  10  & 50 & 50 & \textbf{seems infeasible}   \\ 
 \hline
\end{tabular}
\caption{A list of some set sizes for dimension 4 as well as the conclusion made using our optimisation techniques. Here ``proven feasible'' implies a valid MUSB was found to a tolerance of $10^{-13}$ using method 1, whilst ``seems infeasible'' implies our search for a feasible point using method 1 was inconclusive even after 100000 iterations.}
\label{tbl:resultsSizes4}
\end{table}

\begin{table}[ht]
\centering
\begin{tabular}{| c | c c c c |} 
 \hline
 set sizes & vecs & vars & eqns & result \\
 \hline
 \{5,5,5,5,5,5\} &  30  & 652 & 902 & proven feasible   \\ 
  \hline
\{1,1,1,1,1,1,1\} &  7  & 60 & 40 & proven feasible   \\ 
\{2,2,2,2,2,1,1\} &  12  & 138 & 132 & proven feasible  \\ 
\{2,2,2,2,2,2,1\} &  13  & 162 & 161 & \textbf{seems infeasible}  \\ 
  \hline
\{3,2,2,1,1,1,1\} &  11  & 96 & 90 & proven feasible   \\ 
\{3,3,2,1,1,1,1\} &  12  & 116 & 114 & proven feasible    \\ 
\{3,3,3,1,1,1,1\} &  13  & 136 & 140 & \textbf{seems infeasible}   \\ 
 \hline
\{5,1,1,1,1,1,1\} &  11  & 60 & 55 & proven feasible   \\ 
\{5,2,1,1,1,1,1\} &  12  & 78 & 75 & proven feasible \\ 
\{5,3,1,1,1,1,1\} &  13  & 96 & 97 & \textbf{seems infeasible}   \\ 
 \hline
\end{tabular}
\caption{A list of some set sizes for dimension 5 as well as the conclusion made using our optimisation techniques. Here ``proven feasible'' implies a valid MUSB was found to a tolerance of $10^{-13}$ using method 1, whilst ``seems infeasible'' implies our search for a feasible point using method 1 was inconclusive even after 100000 iterations.}
\label{tbl:resultsSizes5}
\end{table}

\begin{table}[ht]
\centering
\begin{tabular}{| c | c c c c |} 
 \hline
 set sizes & vecs & vars & eqns & result \\
 \hline
 \{6,6,6\} &  18  & 170 & 206 & proven feasible   \\ 
 \hline
\{1,1,1,1\} &  4  & 22 & 7 & proven feasible   \\ 
\{2,1,1,1\} &  5  & 22 & 9 & proven feasible   \\ 
\{2,2,1,1\} &  6  & 36 & 18 & proven feasible   \\ 
\{2,2,2,1\} &  7  & 50 & 29 & proven feasible   \\ 
\{2,2,2,2\} &  8  & 66 & 43 & proven feasible   \\ 
\hline
\{3,2,2,1\} &  8  & 50 & 33 & proven feasible   \\ 
\{3,2,2,2\} &  9  & 66 & 48 & proven feasible   \\ 
\{3,3,3,2\} &  11  & 102 & 86 & proven feasible   \\ 
\{3,3,3,3\} &  12  & 122 & 109 & proven feasible   \\ 
\hline
\{4,2,1,1\} &  8  & 36 & 24 & proven feasible   \\ 
\{4,3,3,3\} &  13  & 122 & 117 & proven feasible   \\ 
\{4,4,3,3\} &  14  & 144 & 144 & \textbf{seems infeasible}   \\ 
\hline
\{6,1,1,1\} &  9  & 22 & 15 & proven feasible   \\ 
\{6,6,1,1\} &  14  & 92 & 95 & proven feasible   \\ 
\{6,6,2,1\} &  15 & 114 & 121 & proven feasible   \\ 
\{6,6,3,1\} &  16  & 136 & 149 & \textbf{seems infeasible}   \\ 
\hline
\{6,3,3,2\} &  14  & 102 & 100 & proven feasible   \\ 
\{6,3,3,3\} &  15  & 122 & 125 & \textbf{seems infeasible}   \\ 
\hline
\{6,3,2,2,1\} &  14  & 106 & 102 & proven feasible   \\ 
\{6,3,2,2,2\} &  15  & 128 & 128 & \textbf{seems infeasible}   \\ 
\hline
\{6,2,2,1,1,1,1,1\} &  15 & 134 & 131 & proven feasible   \\ 
\{6,2,2,2,1,1,1,1\} &  16 & 158 & 160 & \textbf{seems infeasible}   \\ 
 \hline
\end{tabular}
\caption{A list of some set sizes for dimension 6 as well as the conclusion made using our optimisation techniques. Here ``proven feasible'' implies a valid MUSB was found to a tolerance of $10^{-13}$ using method 1, whilst ``seems infeasible'' implies our search for a feasible point using method 1 was inconclusive even after 100000 iterations.}
\label{tbl:resultsSizes6}
\end{table}

The algorithms used here are general and can be applied to any commuting polynomial optimisation problem using our library, written as a single C++ header file. The library represents polynomials using hash tables, allowing for high performance even with large equations. We release this code as an open-source project \cite{mygithub}, although it still needs work regarding user-friendliness and documentation, for now only serving as a reference for the methods used as well as containing the data files for each optimisation referenced. The example relevant to this work can be compiled using ``make mub'' and then run the ``mub'' executable as ``./mub -h'' to display the help text.

\section{Conclusion}

We explored the polynomial representation of the open mutually unbiased bases existence problem, showing a number of reductions in order to take full advantage of the available symmetries. The result is then a second-order real non-linear set of equations. We then show two methods of approaching such a problem, a Lagrange-multiplier Newtonian-descent algorithm which we use for quickly checking which of our relaxations are feasible, as well as a global branch-and-bound algorithm. We introduce the notation of mutually unbiased sub-bases and show that proving infeasibility can be done by proving infeasibility of sub-bases, reducing the number of variables drastically.

Our reductions regarding the MUB problem may simplify future proofs and generate new questions about why sub-bases cannot be found in certain dimensions and why the minimum sub-basis size seems to follow a particular pattern. Unfortunately, solving the open question in this framework by searching the whole space still remains intractable, with our estimates suggesting it is still beyond the capacity of even the most advanced supercomputers at the time of writing. Nevertheless, given humanity's ongoing quest for greater computational power, we are confident that the technology of the near future will bring the ability to solve this problem, accelerated by any further reductions and simplifications we find.

\section{Acknowledgements}

Thanks to Máté Farkas, Leonardo Zambrano and Antonio Acín for the useful discussions.

This  project  has  received  funding  from  the  European  Union’s  Horizon  2020  research and innovation programme under the Marie Skłodowska-Curie grant agreement No 847517.


\pagebreak
\bibliographystyle{unsrt}
\bibliography{refs}

\end{document}